\documentclass[a4paper,fleqn,usenatbib]{mnras}
\usepackage{graphicx}
\usepackage{graphics}
\usepackage{amsmath}
\usepackage{multirow}
\usepackage{color}
\usepackage{xcolor}
\usepackage{bm}		

\usepackage{float}

\def\kms{{\rm\,km\,s^{-1}}}
\def\kmskpc{{\rm\,km\, \,s^{-1} \, {kpc}^{-1}}}

\def\Gyr{{\rm\,Gyr}}

\def\mathnew{\mathsurround=0pt}   
\def\simov#1#2{\lower .5pt\vbox{\baselineskip0pt  
    \lineskip-.5pt\ialign{$\mathnew#1\hfil##\hfil$\crcr#2\crcr\sim\crcr}}}

\def\'#1{\ifx#1i{\accent"13\i}\else{\accent"13#1}\fi}

\def\aap{{A\&A}}

\def\araa{{ARA\&A}}
\def\aj{{AJ}}
\def\apj{{ApJ}}
\def\apjl{{ApJL}}

\def\mnras{{MNRAS}}
\def\apjs{{ApJS}}

\def\rmxaa{{Rev. Mex. Astron. Astrofis.}}
\def\pasp{{PASP}}
\def\msais{{MSAIS}}

\title[Revealing the spiral arms through radial migration and the
  shape of the MDF]{Revealing the spiral arms through radial migration
  and the shape of the Metallicity Distribution Function}
\author[Martinez-Medina et al. 2016]{L.A. Martinez-Medina
  \thanks{Contact e-mail:
    \href{mailto:lamartinez@astro.unam.mx}{lamartinez@astro.unam.mx}},
  B. Pichardo, E. Moreno, \& A. Peimbert \\ Instituto de Astronom\'ia,
  Universidad Nacional Aut\'onoma de M\'exico, A.P. 70--264, 04510,
  M\'exico, CDMX, Mexico}

\begin{document}
\label{firstpage}
\pagerange{\pageref{firstpage}--\pageref{lastpage}}
\maketitle

\begin{abstract}
Recent observations show that the Milky Way's metallicity distribution
function (MDF) changes its shape as a function of radius. This new
evidence of radial migration within the stellar disc sets additional
constraints on Galactic models. By performing controlled test particle
simulations in a very detailed, observationally motivated model of the
Milky Way, we demonstrate that, in the inner region of the disc, the
MDF is shaped by the joint action of the bar and spiral arms, while at
outer radii the MDF is mainly shaped by the spiral arms. We show that
the spiral arms are able to imprint their signature in the radial
migration, shaping the MDF in the outskirts of the Galactic disc with
a minimal participation of the bar. Conversely, this work has the
potential to characterise some structural and dynamical parameters of
the spiral arms based on radial migration and the shape of the
MDF. Finally, the resemblance obtained with this approximation to the
MDF curves of the Galaxy as seen by APOGEE, show that a fundamental
factor influencing their shape is the Galactic potential.

\end{abstract}                
 

\section{Introduction} \label{sec:intro}
An enormous effort has been dedicated, in the last decade, to the
understanding of radial and vertical orbital stellar motions induced
by the non-axisymmetric structures in the Galaxy \citep[and references
  therein]{Sellwood2002,2008ApJ...684L..79R,2012MNRAS.421.1529G,2012MNRAS.420..913B,2012MNRAS.426.2089R,VeraCiro2014,HalleDiMatteo2015,2016arXiv160400191A}. Particular
attention has been paid to stellar migration and its effects on the
chemical elements distribution in the Galaxy
\citep{Schonrich2009,Minchev2011,2013A&A...558A...9M,2014A&A...572A..92M,SanchezBlazquez2014,Hayden2015,2016ApJ...818L...6L}.

The mechanism of radial migration, as defined by \citet[hereafter
  SB02]{Sellwood2002}, is understood as the redistribution of angular
momentum for stars that interact with the non-axisymmetric structure
of the galaxy around the corotation resonance while keeping their
orbital eccentricity unchanged. On the other hand, redistribution of
angular momentum at radii different from corotation will cause a
dynamical heating of the stellar disc. Both mechanisms move stars to
inner or outer radii, but radial migration does not leave a kinematic
imprint on stellar orbits. It is worth noting that the term radial
migration has been used differently by different authors \footnote{The
  results presented in this paper do not depend on the specific
  definition of radial migration.}  \citep[see for example the
  discussion by][]{VeraCiro2014}.

Without the effect of stellar migration, a perfect correlation between
age and metallicity of a star could be found for a given Galactic
region, assuming the abundance of chemical elements was known in such
region. Although this seems to approximate the case for the ISM in the
Milky Way (MW) and other galaxies
\citep{1994ARA&A..32..191W,1999PASP..111..919H}, it is known that, for
example, in the solar neighbourhood, stars of a given age show a large
dispersion in metallicity \citep{1993A&A...275..101E}. This effect is
not readily explained by plain orbital excursions from epicycles
corresponding at their birth place (SB02).

More recently, \citet{Hayden2015}, measure the metallicity
distribution functions (MDFs) of the MW, from a sample of 69,919 red
giants from the SDSS-III/APOGEE Data Release 12. They find that the
shape of the midplane MDF changes systematically with radius, with a
negatively skewed distribution at 3 $<$ R $<$ 7 kpc, to a roughly
Gaussian distribution at the solar annulus, to a positively skewed
shape in the outer Galaxy. Using a simple model they suggest that the
reversal of MDF shape could be due to radial migration. However, a
more complex model is needed to differentiate between the contribution
of the different non-axisymmetric components of the Galaxy.

With detailed orbital studies performed in suitable observationally
motivated potentials of the MW, we show here that the spiral arms can
imprint their mark on the MDF.

This paper is organised as follows. The galactic model, initial
conditions, and methodology are described in Section \ref{model}.  A
study on radial migration and radial heating is presented on Section
\ref{Results}. The link between the MDF and the initial radial
distribution is shown in Section \ref{Linking}. Finally we present the
discussion and conclusions, in Sections \ref{discussion} and
\ref{conclusions}

\section{The Galaxy model, numerical simulations and initial conditions}
\label{model}
A good numerical approach to study radial migration and its relation
to MDFs in a Milky-Way sized N-body simulation was recently provided
by \citet{2016ApJ...818L...6L}. N-body simulations are not suitable to
achieve the goals of this work, as we explain below. We employ instead
a very detailed steady model adjusted specifically to the MW to the
best of recent knowledge of the Galaxy structural and dynamical
parameters (i.e spiral arms and bar masses, density laws,
scale-lengths, angular velocities, etc.).

We have selected this elaborate but steady model over a sophisticated
N-body simulation (without resorting to the more common and simplistic
cosine potential for the spiral arms with a Ferrers bar) for the
following reasons:

1) The model is fully adjustable. We are able to fit the whole
axisymmetric and non-axisymmetric potential (i.e. spiral arms and
bar), in three dimensions to our best understanding of the Milky Way
(or any other particular galaxy) from observations and models.

2) Rather than using a simple ad hoc model for a spiral perturbation,
we employ a three-dimensional (3D) mass distribution for the spiral
arms, from which we derive their gravitational potential and force
fields. Our model is considerably faster, computationally speaking,
than N-body simulations.

3) It allows us to study in great detail individual stellar orbital
behaviour (e.g. resonant regions, vertical structure, chaotic and
ordered behaviour, periodic orbits to estimate at some degree orbital
self-consistency, etc.), without the resolution problems of N-body
simulations.

We have integrated test particle orbits in this 3D Galactic potential
model. Our model is observationally motivated by the MW and suitable
to explain several characteristics of the local kinematics due to the
spiral arms and the bar (such as moving groups in the solar
neighbourhood, e.g. \citet{2009ApJ...700L..78A}).

Although the one employed is a much better suited spiral model to
represent the Milky Way than any N-body simulation, we do not include
any prescription of metallicity or ISM physics as in
\citet{2016ApJ...818L...6L}. In a future work, we will perform a more
specific study of the spiral arms parameters, as well as implementing
a metallicity prescription, to seek for a better fit to the APOGEE
MDFs for the Milky Way Galaxy, and search for some restrictions to the
morphology and physical characteristics of the spiral arms.

\subsection{The full Galactic model}
\label{armbar}
\begin{figure}
\begin{center}
\includegraphics[width=9cm]{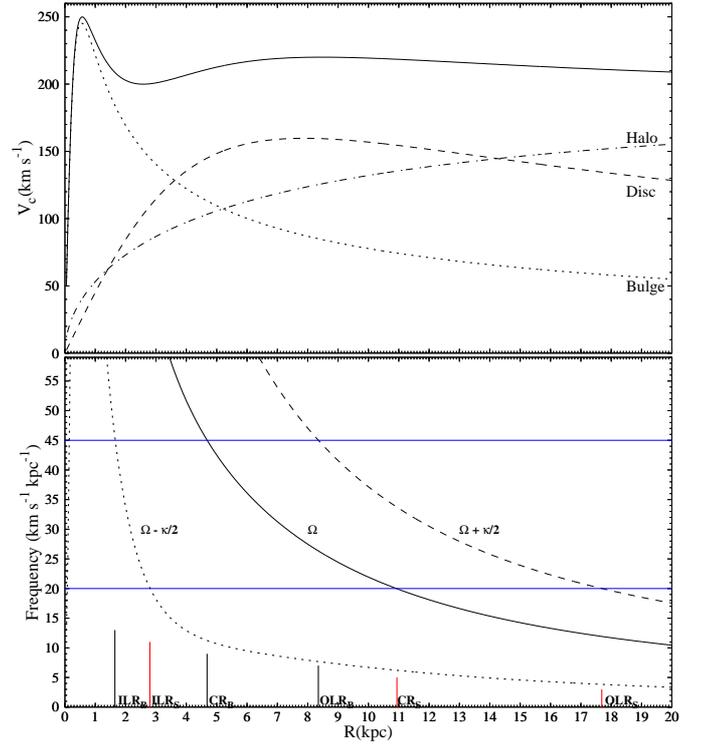}
\end{center}
\caption{Top: Circular velocity for the bulge, disc, halo, and
    the full axisymmetric component of the model.  Bottom: Frequencies
    diagram of the model. The vertical lines indicate the position of
    the inner and outer Lindblad resonances (ILR, OLR) and corotation
    (CR) for both the bar (black) and spiral arms (red). The
    horizontal blue lines represent the pattern speed of the bar
    (45$\kmskpc$) and the spiral arms (20$\kmskpc$).}
\label{fig:rcyres}
\end{figure}

The Galactic model is based on an axisymmetric model with the addition
of non-axisymmetric components (spiral arms and bar). The axisymmetric
galactic potential is based on the potential of \citet{AS91}, that
consists of a disc, a spherical bulge, and a supermassive spherical
halo. All the mass of the spherical bulge is used to build the
Galactic bar (see Section \ref{bars}), likewise, the 3D spiral arms
(see Section \ref{arms}) are introduced by reducing the mass of the
disc. At the beginning of each run, both structures are introduced
slowly to diminish transients, by increasing their masses, as the
bulge and disc masses are reduced. Finally, the Galactic potential is
scaled to the Sun's galactocentric distance, $R_0$ = 8.5 kpc, and the
local rotation velocity, ${\Theta}_0$ = 220 km s$^{-1}$. The rotation
curve and frequencies diagram with the correspondent resonances are
indicated in Figure \ref{fig:rcyres}.

\subsection{The spiral arms}
\label{arms}
Numerous observational papers based on the younger components of the
Galactic disc (HII regions, O-B stars, CO emission, masers in
high-mass star-forming regions) show a four-armed spiral large scale
structure in the Galaxy. On the other hand, observations in infrared
bands such as those of the COBE/DIRBE K-band and the infrared
Spitzer/GLIMPSE survey, show that only two of the observed arms seem
dominant \citep{2001ApJ...556..181D,CHBet09}. Based on models that
have shown that two additional gaseous arms can be formed (without
increasing the stellar surface density) as a response to a two-armed
dominant pattern \citep{2004MNRAS.350L..47M}, in this work we will
adopt a two-armed structure for the spiral arms.

For the spiral arms we employ the PERLAS model of \citet{PMME03}. The
model is formed by a two-armed 3D density distribution made of
individual inhomogeneous oblate spheroids. The spheroids act as bricks
in a building, to construct the arms structure; they are located along
a logarithmic spiral locus. PERLAS is completely adjustable (i.e. the
width, height, scale lengths, density fall along the spiral arms and
transversal to them, etc.) to better represent the available
observations.

Our spiral arms simulate the main Galactic spiral arms based on the
Spitzer/GLIMPSE database \citep{BCHet05,CHBet09}. The density is
distributed as an exponential decline along the arms. The total mass
of the spiral arms taken in these experiments is 4.28 $\times$ 10$^9$
M$_{\odot}$, that corresponds to a mass ratio of $M_{\rm arms}/M_{\rm
  disc}$ = 0.05. 

We employ the function $Q_T$ \citep{ST80,CS81} to measure the strength
of the arms (which is proportional to their total mass). The maximum
value of the function $Q_T$ over the radial extent of the spiral arms
(known as $Q_s$) is calculated. The value obtained for $Q_s$ is
smaller than 0.25, which is in agreement with \citet{BVSL05}, that
finds that, for a galaxy like the Milky Way (Sbc), $Q_s$ is
approximately 0.25.

Finally, for the angular velocity, we use a value of ${\Omega}_S = 20
\kmskpc$, provided by different observational and theoretical methods
(see \citet{G11} for a review).

\subsubsection{Transient spiral arms}
\label{trans}
There is no consensus in the astronomical community on the nature of
spiral arms: fixed grand design vs transient and dynamic
spirals. While some competent N-body simulations have shown
long-lasting spiral arms \citep[e.g.][]{2016arXiv160701953S}, the most
of them seem to show spiral arms as structures of transient nature
\citep{1974MNRAS.167..351H,1984ApJ...282...61S,2008ApJ...684L..79R,2011MNRAS.410.1391A,2014ApJ...785..137S}.

We have made an effort to consider the likely possibility that
  real galaxies have transient spiral arms. To this purpose, we
  introduce gradual changes in amplitude and angular speed in the
  arms. Note that, although we are not changing the pitch angle
  directly, by changing the amplitude we can represent changes not
  only to the mass but also mimic the effect of changing the pitch
  angle, which itself directly impacts the azimuthal forcing.

The construction of transient arms uses a simple assumption: the
amplitude (equivalently the mass) of the spiral arms changes with a
given periodicity. In this way the mass of the arms in the model can
be made to depend explicitly on time, for which we have chosen the
following time dependence:

\begin{equation}
\label{eq:transient}
M_s(t) = M_{max}\mid\sin(\pi t/t_l)\mid^f, 
\end{equation}
where $M_s$ is the mass of the spiral pattern, $t_l$ is the lifetime
of each pattern (set to 1 Gyr), $M_{max}$ is the maximum value of the
arms' mass during its lifetime, and the parameter $f$ (set to 0.5)
controls the flatness of each ripple in a plot of $M_s(t)$, i.e., it
controls how long the mass remains nearly quiescent, and how fast
$M_s$ grows and diminishes outside the quiescent period. Similarly, to
mimic a changing pattern speed we implemented a time dependent
rotation frequency $\Omega(t)$. Although not conclusive, it
  seems from some N-body simulations, that new born spiral arms tend
  to decrease their angular speeds \citep{2011MNRAS.410.1637S}. As a
first approximation we set the rotation frequency of the initial
pattern in the simulation to $25\kmskpc$ and decreased it with each
new pattern, to a minimum of $17\kmskpc$, at the end of the
simulation. Since a functional form of $\Omega(t)$ is not evident from
N-body simulations, a simple assumption is that $\Omega$ is constant
during each time period $t_l$, in this way the adopted funcional
expression for $\Omega(t)$ is a uniformly decreasing step function,
each step indicates the value of the pattern speed for each
short-lived spiral. We are aware that there is evidence of a
  more general relation for the pattern speed in N-body/dynamic
  spirals, that traces the material rotation frequency
  \citep[See][]{1999A&A...348..737R,2013ApJ...763...46B,2013MNRAS.432.2878R,2014MNRAS.444.3756M,2015MNRAS.449.3911P,2012MNRAS.426..167G};
  in a future work we will study the dynamical effects of this
  specific characteristic of the spiral arms implemented on
  potential-particle models of the Galaxy.

As stated above, with the transiency implemented in this way, the main
goal is not to reproduce all the features of dynamical spirals, but to
test the impact of a time changing amplitude and pattern speed on the
mechanism of radial migration, which is predicted be sensitive to both
features.

\subsection{The bar}
\label{bars}
For the bar we have selected the triaxial bar model of
\citet{Pichardo2004}; this is an inhomogeneous ellipsoid built as a
the superposition of a large number of homogeneous ellipsoids to
achieve a smooth density fall. The model approximates the density fall
fitted by \citet{F98} from the COBE/DIRBE observations of the Galactic
centre. The total mass of the bar is 1.4 $\times$ 10$^{10}$
M$_{\odot}$, within the observational limits
\citep[e.g.,][]{K92,Z94,DAH95,B95,SUet97,WS99,Aet14}. Regarding the
angular speed, a long list of studies have estimated this parameter
\citep[and references therein]{G11}, concluding that the most likely
value lies in the range ${\Omega}_B=$ 45 -- 60 $\kmskpc$. For our
computations, we adopt the value ${\Omega}_B$ = 45 $\kmskpc$, based on
the formation of moving groups in the solar neighbourhood
\citep{2009ApJ...700L..78A}. 

A brief description of the observational/theoretical parameters
utilised to model the bar and spiral arms is summarised in Table
\ref{tab:model}. For more specific details on the construction and our
fit to observations see
\citet{2015ApJ...802..109M,Pichardo2004,PMME03,2015MNRAS.451..705M,2012AJ....143...73P}.

\begin{table*}
\begin{minipage}{90mm}
\caption{Parameters of the non-axisymmetric Galactic components}
\label{tab:model}
\begin{tabular}{lcr}
 \hline
 \hline

Parameter & Value & Reference\\
 \hline

\multicolumn{3}{c}{\it Triaxial ellipsoidal Bar}\\
 \hline
Major Semi-Axis                     & 3.5 kpc            & 1 \\
Scale Lengths                       & 1.7, 0.64, 0.44 kpc & 1 \\
Axial Ratios                        & 0.64/1.7, 0.44/1.7  & 1  \\
Mass                                & 1.4 $\times$ 10$^{10}$ M$_{\odot}$     &  2,3,4,5,6 \\
Pattern Speed ($\Omega_B$)          & 45,20 $\kmskpc$  & 7,8 \\
 \hline
\multicolumn{3}{c}{\it Spiral Arms}\\
 \hline
Number of spiral arms                   & 2 &  9,10 \\
Pitch Angle ($i$)                   & 15.5$^{\circ}$ & 11 \\
Radial Scale Length ($H_{\star}$)          & 3 kpc        & 12 \\
$M_{\rm arms}/M_{\rm disc}$         &  0.05         & 13 \\
Mass                                &  4.28 $\times$ 10$^9$ M$_{\odot}$           & 13 \\
Pattern Speed ($\Omega_S$)          & 20 $\kmskpc$     & 8 \\
 \hline
$R_0$                               & 8.5 kpc       &  \\
 \hline
\end{tabular}

{           (1)~\citet{F98};
            (2)~\citet{G02};
            (3)~\citet{2008A&A...480..723C};
            (4)~\citet{1996MNRAS.283..149Z};
            (5)~\citet{B95};
            (6)~\citet{DAH95};
            (7)~\citet{Aet14};
            (8)~\citet{G11};
            (9)~\citet{2001ApJ...556..181D};
            (10)~\citet{CHBet09};
            (11)~\citet{D00};
            (12)~\citet{BCHet05};
            (13)~\citet{PMME03}.}

\end{minipage}
\end{table*}

\subsection{Initial conditions}
\label{armbar}The initial distribution of test particles follows a 
Miyamoto-Nagai density profile discretised in $10^6$ particles; the
particles are distributed in the velocity space following the strategy
of \citet{Hernquist1993}. When tested, the velocity dispersion does
not evolve in the presence of the axisymmetric part of the model. Once
the initial particle disc is constructed, the bar and spiral arms grow
adiabatically for a time $t_0$, after which the mass of each component
remains fixed. We follow a smooth function to grow the
non-axisymmetric structures provided in equation 4 of
\citet{Dehnen2000} that avoids any sudden relaxation of the disc.

The mass of the spiral arms is given as a fraction of the mass of the
disc and the mass of the bar is taken from the mass of the
axisymmetric bulge. This means that, starting with the axisymmetric
components of the model at time $t = 0$, during the time interval $0 <
t < t_0$, a fraction of the disc mass is transferred to the arms and
all the mass of the bulge is transferred to the bar.

A thorough test of the disc heating, measured from radial and vertical
velocity dispersions, led us to take $t_0 = 0.5$ Gyr as enough time to
guarantee the absence of artificial effects in the simulations
(i.e. heating).

The transient spiral arms are introduced adiabatically, as we do with
the bar and the steady arms; the growing time for these experiments is
0.5 Gyr. Once the arms reach their maximum mass, the functional form of
equation \ref{eq:transient} will take place to mimic transiency by
decreasing and increasing the mass of the spiral pattern. Notice that
this functional dependence is not as smooth as our adiabatic growth
(with this function we are trying to reproduce the behaviour seen, in
general, in N-body simulations).

\section{Radial migration and radial heating}
\label{Results}

The first aim of this work is to isolate the contributions of the
spiral arms and of the central bar, in the context of stellar radial
migration, and then to compare them to the full Galactic model.

\subsection{Radial migration by the spiral arms}
\label{Arms}
For the first MW model, the only non-axisymmetric component we
consider is that of the spiral arms. We measure radial migration and
heating in terms of changes in angular momentum calculated for each
particle in the simulation during the period of time, $\Delta t = t_f
- t_i = 1$ Gyr, where the initial time, $t_i=4$ Gyr, and the final
time, $t_f=5$ Gyr.

\begin{figure*}
\begin{center}
 \includegraphics[width=18cm]{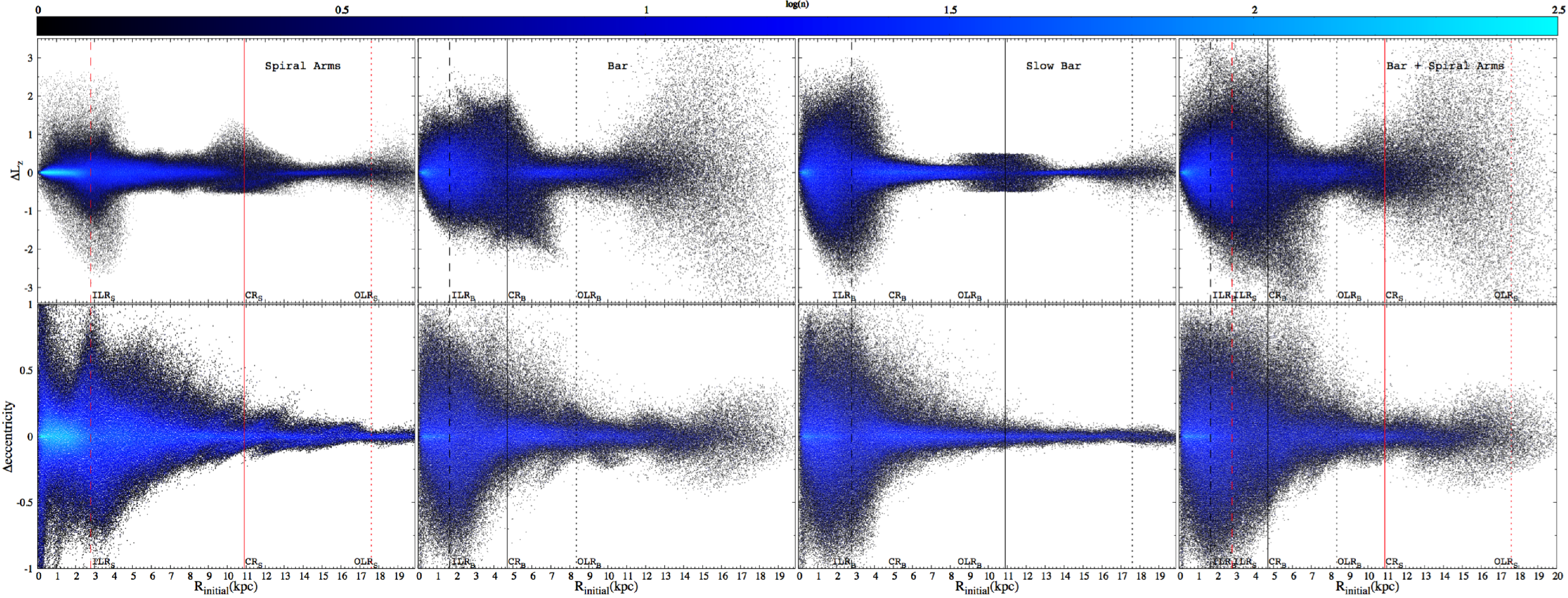}
\end{center}
\caption{Changes in angular momentum (top row) and changes in
  eccentricity (bottom row), both as a function of the initial radius
  $R_i$. First column: simulation with spiral arms + axisymmetric
  potential. Second column: simulation of the bar + the axisymmetric
  potential. Third column: slow-bar + the axisymmetric
  potential. Fourth column: simulation of the full model (spiral arms
  + Galactic bar + axisymmetric potential). $\Delta L_z$ is in units
  of kpc $\times v_c$, with $v_c = 220 \kms$. The vertical lines mark
  the positions of the ILR (dashed), CR (solid), and the OLR (dotted)
  of the spiral pattern (red) and the bar (black). The color map
  indicates the number of particles $log(n)$.}
\label{fig:dL}
\end{figure*}

In the upper left panel of Figure \ref{fig:dL}, we present the change
in angular momentum, $\Delta L_z$, for each particle as a function of
the initial radii, $R_i$, in the time interval, $\Delta t$. The
resonance locations of the spiral pattern are indicated in the
figure. Note that all clear changes in angular momentum are linked to
resonances. The theory predicts that changes that occur around the
inner and outer Lindblad resonances are related to radial heating of
the stellar disc, while $\Delta L_z$ occurring at corotation, are
mainly related to radial migration (SB02).

To test this prediction, in the lower left panel of Figure
\ref{fig:dL} we present the changes in orbital eccentricity within the
same period of time, $\Delta t$, as a function of the initial
radii. It can be seen that the gains and losses of angular momentum
around the ILR correspond to changes in the orbital eccentricity
(i.e. radial heating). On the other hand the gains and losses of
angular momentum around CR are similar in magnitude to what happens
around the ILR, but show a very different kinematic behaviour: in
corotation the changes in the orbital eccentricity are much smaller
than in the ILR, indicating the presence of radial migration.

\subsection{Radial migration by the bar}
\label{Bar}
We explore now the isolated effect of the bar in the MW-like
model. Proceeding as in the previous case, we measure $\Delta L_z$ as
a function of the initial radii for each particle within a period of
time of $\Delta t = 1$ Gyr. As shown in the upper second column panel
of Figure \ref{fig:dL}, these changes occur along the stellar disc,
linked to the bar resonances, as with the spiral pattern case; in this
scenario, the distribution of $\Delta L_z$ {\it vs} $R_i$ is different
from the spiral arms case, first, because the bar is a more massive
structure, but mostly because of the greater pattern speed.

Although the ILR and CR of the bar are quite close, the distributions
of $\Delta L_z$ associated with each of them are still
distinguishable. Note how particles outside OLR spread out, a similar
but larger effect than the one seen with the spiral arms. This effect
is mainly heating due to the OLR in both the bar and the spiral arms
cases, as predicted by SB02. Note also that in the spiral arms case,
the scatter in $\Delta L_z$ develops around and beyond the OLR (as
predicted theoretically), but in the bar case, the scatter starts a
little beyond the OLR. This might be due to the diminishing surface
gravity of the disc with the galactocentric distance. However, from
the color map in Figure \ref{fig:dL}, it can be seen that the number
of scattered particles is actually very small; a deeper study of this
effect will be presented in a future work.

Again, by measuring the changes in orbital eccentricity, we can
establish which part of the diagram $\Delta L_z$ vs $R_i$ represents
radial heating or radial migration. From Figure \ref{fig:dL} (second
column, bottom panel), it is clear that around the ILR and beyond the
OLR radial heating predominates because at these regions the stars
modify their angular momentum by changing their orbital
eccentricity. On the other hand, note how changes in eccentricity
diminish as moving from the ILR to CR; the bimodality in the $\Delta
L_z$ distribution around CR of the bar implies that stars can change
their radii by several kiloparsecs without heating their orbits.

\subsubsection{Slower Bar}
As mentioned in the previous section, changes in angular momentum
occur mainly around the resonances of the bar. As in the previous
case, $\Delta L_z$ around the ILR heats the orbits by changing
considerably its eccentricity. On the other hand, $\Delta L_z$ around
CR is quite large but with small changes in the orbital eccentricities
around that region. These two different outcomes for similar changes
in angular momentum allow us to distinguish between radial heating and
radial migration.

However, although a pattern speed of $\Omega = 45$ km s$^{-1}$
kpc$^{-1}$ is a common estimate for the MW bar, this value sets CR in
close proximity to the ILR and it is not very clear what happens
around CR. As an illustrative example we ran a simulation with a
slower bar: $\Omega = 20$ km s$^{-1}$ kpc$^{-1}$ (the same used for
the spiral arms).

Figure \ref{fig:dL} shows also the two simulations with the bar-only
model: one with a pattern speed of $\Omega = 45$ km s$^{-1}$
kpc$^{-1}$ (second column), the other with $\Omega = 20$ km s$^{-1}$
kpc$^{-1}$ (third column). First, notice that with a smaller pattern
speed the separation between the ILR and CR increases; consequently,
it is clearer what changes in angular momentum are linked to a given
resonance. Around the ILR, the slow bar exchanges angular momentum
with a substantial amount of heating, as seen by the changes in
eccentricity. Meanwhile, there is a distinctive feature, around CR, in
$\Delta L_z$ of the slow bar with small changes in orbital
eccentricity, as expected for radial migration; however, these changes
are much smaller than those around CR of the faster bar, they even are
smaller than the ones in the spiral-only model with the same pattern
speed.  This is mainly because, although the exchange of angular
momentum is linked to resonances, in particular to CR, the bar (or
spiral) should reach physically that given radius in order to
accelerate the particles towards itself. Therefore, a faster bar, with
CR located at a small radius, is efficient exchanging angular momentum
with the stellar disc through radial migration when compared with a
slow bar, for which CR is well outside.

\subsection{The Full Galactic Model (bar + spiral arms)}
\label{Bar+Arms}
By isolating the effect of the spiral arms and the bar, we were able
to better distinguish radial heating and radial migration induced by
each non-axisymmetric structure, as well as their zones of influence
within the disc. Now, the test particle disc is evolved in a full MW
galactic model that includes both a bar and spiral arms. Top right
panel of Figure \ref{fig:dL} shows $\Delta L_z$ in a period of time of
$\Delta t = 1$ Gyr.

Compared to the previous cases $\Delta L_z$ is larger in magnitude and
more spread throughout the disc. The resonance positions are indicated
in Figure \ref{fig:dL}; at inner radii there is an overlapping of
resonant regions, specifically the ILR of the bar and the ILR of the
arms, which enhances the gains and losses of angular momentum when
compared with the individual effect of either one.

With this study it is possible to elucidate whether resonance
overlapping preserves the eccentricity and induces radial migration
\citep{Minchev2010,Minchev2011}, or instead it only heats up the
disc. Figure \ref{fig:dL} (bottom right panel), shows the changes in
orbital eccentricity in the period of time $\Delta t$. Note that
within the region of resonance overlapping the diagram is dominated by
substantial changes in eccentricity, i.e., the redistribution of
angular momentum at this zone is not produced by radial migration, but
preferentially by radial heating. This occurs because the resonances
in that region are the ILR of the bar and the ILR of the arms, each of
which is related to radial heating. Additionally, since these
resonances are close to the CR of the bar, they reduce the radial
migration that could be induced by CR.

On the other hand, the CR of the spiral pattern is located at a large
radius where no resonance overlapping is present; consequently, the
redistribution of angular momentum around the CR of the arms (as seen
in top left panel of Figure \ref{fig:dL}) is preserved, even in the
presence of the bar, and it can still be classified as radial
migration because changes in orbital eccentricity are small.

Concerning the model with bar + transient spiral arms, we found that
the changes in angular momentum are similar in magnitude to the ones
in the model with steady arms, however those pervade on a large region
(several kiloparsecs) of the disc; it is of little interest to present
the same plot as Figure \ref{fig:dL}, since for the same period of
time it would represent arms with constant speed, and when we plot the
full 5 Gyr it just looks like a blurred image of Figure
\ref{fig:dL}. In the next section we plot the initial radii
distributions along the stellar disc, which are more suitable to study
the cumulative effect on radial migration by both kinds of
  spiral arms (transient and steady).

Summarising, the bar and spiral arms, separately, induce radial
heating and radial migration, and their zones of influence over the
disc can be determined. When both structures are present: at inner
radii, the radial mixing of the stellar disc is produced by the joint
action of the perturbers and it is dominated by radial heating; while
at outer radii, the redistribution of angular momentum is caused
mostly by the spiral arms and dominated by radial migration. This
leads to an important conclusion: the imprint of the spiral arms may
be identified even in the presence of the bar.

\section{Linking the MDF to the initial radii distribution}
\label{Linking}
Using the results of Section \ref{Results}, we will now demonstrate
that the spiral arms are able to present a characteristic signature of
their existence in observations.

\subsection{Initial radii distribution modelled with the different Galactic components}
By comparing the experiments presented previously we can identify, from a
theoretical point of view, the effect that the spiral arms and/or the bar
drive into the radial mixing of the stellar disc, whether this is done
by means of radial heating or radial migration, and the regions where
it takes place.

Observationally speaking, it is not plausible to identify the radial
migration occurring in the Galaxy trough plain kinematics of stars. To do
this it is necessary to add chemical labels to the stars.
Former work on radial migration, including chemical evolution of the
disc, predicts that a consequence of this mechanism is a flattening of
the metallicity gradient \citep{Schonrich2009}. However, this
prediction might not be enough to confirm the presence of radial
migration in the stellar disc \citep{SanchezBlazquez2014}.

Using a sample of red giants from the SDSS-III/APOGEE Data Release 12,
in their Figure 5, \citet{Hayden2015}, discovered new signatures of
radial migration by measuring the MDF of stars from inner to outer
radii in the MW; they found a change of shape of the MDF: a negatively
skewed distribution at inner radii that changes to be positively
skewed at large galactocentric radii. Using a simple model that
includes radial migration, \citet{Hayden2015} are able to reproduce
the observed MDF. This observed behaviour, and the interpretation of
the change in shape of the MDF as radial migration, was later found in
N-Body+SPH simulations by \citet{2016ApJ...818L...6L}.

Taking a different approach, employing test particle simulations
within a controlled observationally motivated model of the Galaxy, we
find the same evidence for radial migration, but as we demonstrated in
the sections above, our approach has the advantage of being able to
isolate the effect of each perturber. In this manner, we endeavour to
distinguish, at each Galactic radius, whether the shape of the MDF is
due to the bar, the spiral arms or both.

Since a star preserves information of the state of the ISM at its
birth place and epoch, the trend in the MDF skewness due to radial
migration is reflected directly on the initial radii
distribution. Therefore, instead of working with the MDF, we
equivalently will employ the initial radii distribution of stars in
our simulations, as a tracer of chemical abundance (note that small
radii correspond to large chemical abundance in our scheme).

\begin{figure}
\begin{center}
\includegraphics[width=9cm]{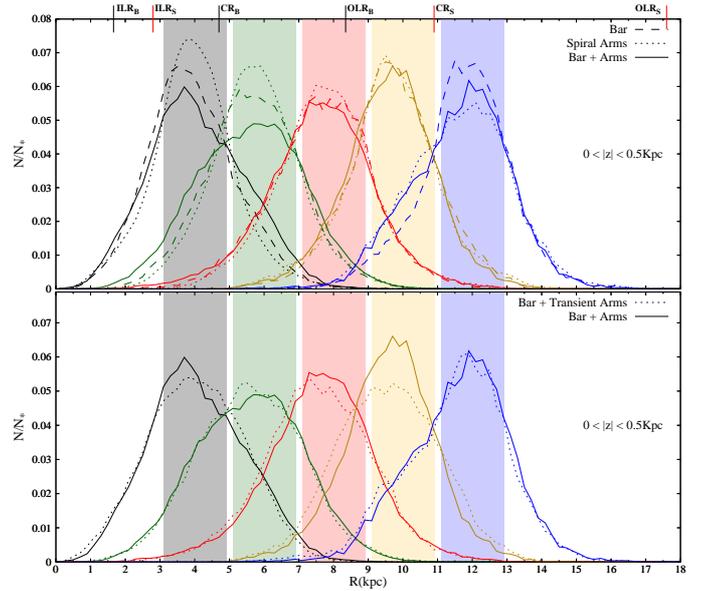}
\end{center}
\caption{Initial galactocentric radii distribution of stars that at
  the end of the simulation are located within one of the five
  coloured bins. Each curve corresponds to the closest same-colour
  shaded bin. $N$ is the number of stars at a given initial radius
  and $N_*$ is the total number of stars that have ended within each
  coloured bin.  Top: distributions for the three models; Spiral Arms
  (dotted), Bar (dashed), and Bar+Arms (solid). Vertical lines at top
  mark the positions of the ILR, CR, and the OLR for the spiral pattern
  and the bar.  Bottom: distributions for the models Bar + long-lived
  Arms (solid) and Bar + transient Arms (dotted).}
\label{fig:Distribution1}
\end{figure}

Figure \ref{fig:Distribution1} shows the initial radii distributions,
i.e., the number of stars with a given initial galactocentric radius
(at 0 Gyr) that at the end of the simulation (at 5 Gyr) are located
within any of the different coloured bins, between $3<$ R $<$ 13
kpc. For purposes of comparison, in the top panel we plot together the
distributions obtained with the three Galactic models that we have
used so far. In the three models the skewness of the distribution is
reversed when going from inner to outer galactocentric radii, as
expected from radial migration, but the asymmetry is more pronounced
with the model that holds both bar and spiral arms.

Complementary to the conclusions of Section \ref{Bar+Arms}, the
comparison of the initial radii distribution, showed in top panel of
Figure \ref{fig:Distribution1}, offers a new method to diagnose the
perturber that induces the radial displacements of stars at each
radius.

First note that, at the inner disc, the initial radii distribution
produced by each perturber does not resemble that of the full model,
meaning that both are important in shaping the MDF in this part of the
Galaxy. Around the solar radius the distributions produced by the
three models are quite similar, therefore one could be inclined to say
it is impossible to determine which one is more influential, but with
the precision achieved with this kind of models one can see that for
the full model the inner tail of the curve is closer to that found
when only the bar is considered, showing that in this region the
effect of the bar is more important than that of the spiral
arms. Finally at large galactocentric radii, there is a change in the
skewness of the distribution, making the effect of the full model much
more similar to the model with only the spiral arms. This means that
at large radii, the MDF of the disc would be mainly shaped by the
radial migration driven by the arms, with little influence from the
bar.

Differences in the shapes of the tails for different models can
potentially be observed with a large enough dataset, allowing to set
restrictions for the structural and/or dynamical parameters of the
non-axisymmetric structures of the MW.

The time interval to compute the diagram in Figure
\ref{fig:Distribution1} is $0-5\Gyr$.  To ensure that neither the
distinction between models nor the sign of the skewness for the
distributions (in Figure \ref{fig:Distribution1}) is due to
short-period transient features, in Appendix \ref{app:TimeFrames}, we
computed the initial radii distributions for two additional time
frames. There we find that the models are distinguishable
independently of the chosen time-frame. This strengths the
significance of the differences between models as well as our main
conclusion: that the spiral arms can leave a distinguishable imprint
on the MDF, even in the presence of the bar.

\subsection{Full model with transient spiral arms}
Radial stellar migration is expected to be more efficient in the
presence of transient spiral patterns (SB02).  We have also
implemented the full Galactic model with spiral arms growing and
vanishing in a transient and smooth fashion, emerging each time with a
decreasing pattern speed, going from 25 down to 17 km s$^{-1}$
kpc$^{-1}$ (for a complete description of the transient spiral arms
model, see Section \ref{trans}).

The bottom panel of Figure \ref{fig:Distribution1} shows the initial
radii distributions for the simulation of the full model (Bar +
long-lived arms) compared to the model of Bar + Transient arms. Note
that as a consequence of the changing pattern speed, the distributions
are modified at all radii when comparing with the case of long-lived
arms. Also the peak of the distribution around the solar vicinity is
slightly displaced towards a smaller radius, this is translated to a
solar neighbourhood MDF with a peak near or slightly over solar
metallicity.

Meanwhile at outer radii the number of stars that migrate from the
inner disc to the two outer bins is increased in the case of the bar +
transient arms model; i.e., a transient spiral pattern would take more
metal rich stars to the outer disc, causing in this region a more
extended high-metallicity MDF tail, populated by stars that were born
in the inner Galaxy.

The shape of the initial radii distributions induced by transient
spiral arms, and its consequences to the MDF, are in good agreement
with the APOGEE observations \citep{Hayden2015}. In fact these
observations, that provide new evidence for radial migration, could be
used to set constraints on chemo-dynamical Galactic models, through
this we could even obtain an insight of the long-lived or transient
nature of the spiral arms, or of the structural parameters of the
spiral arms in a given galaxy, particularly in the MW (in a future
work we will present a related study).

\section{Discussion} 
\label{discussion}
Through controlled test particle simulations in an observationally
motivated detailed model of the MW, we isolate and compare the radial
mixing in three different models: axisymmetric $+$ spiral,
axisymmetric $+$ bar, axisymmetric $+$ spiral + bar (full model). With
this study we are able to distinguish whether a region of the disc is
affected by either of the perturbers or by its joint action.

For the full model, the kinematics that arms and bar imprint to the
stars suggest that at inner radii (less than approximately 5 kpc for
the angular speeds in our model) the radial mixing of the stellar disc
is dominated by the joint action of the perturbers and it is
constituted mainly by radial heating. At outer radii (beyond
approximately 9 kpc) the redistribution of angular momentum is driven
mainly by the spiral arms and dominated by radial migration. The exact
details of the influence region of each structure depends on the
structural and dynamical details of the model such as, masses, angular
velocities, etc.

Radial migration leaves an imprint on the MDFs of the stellar disc;
also transient spiral arms slightly enhance radial mixing and the
skewness of the MDF when compared with long-lasting spiral arms; the
exact shape of the MDF depends on the specific characteristics of the
Galaxy. Comparison of the results of models with the DR12 from APOGEE
can place new constraints on the details of structural and dynamical
parameters of the spiral arms.

The MDF in the inner region of the disc is shaped by the action of
both bar and spiral arms, while at outer radii the MDF is mainly
shaped by the spiral arms. With this work we demonstrate that the
spiral arms show an specific signature readily identifiable on the
MDFs. More work is needed to fully fit the APOGEE MDFs.

Our approach does not consider the chemical abundance nor its
evolution, instead we are employing the initial galactic position as a
proxy for metallicity. However, the similarities obtained between our
models and the observed MDF curves of the Milky Way Galaxy, show the
relevance of the galactic potential in shaping the MDFs curves;
additionally, the difference in the MDFs between the three models
presented here prove that the specific potential employed to model the
Galaxy is vital.

\section{Conclusions} 
\label{conclusions}
With the use of a very detailed fully adjustable potencial, to better
represent the Milky Way Galaxy ---to the best of our observational
knowledge---, we performed simulations to study radial migration and
to approximate the metallicity distribution function to elucidate the
importance of the non-axisymmetric components of the Galaxy (bar and
spiral arms). With this model we are able to split the contribution of
different galactic components to a given phenomenology.

We are able to separate the effect of the bar and spiral arms on
radial heating and radial migration as well as their zones of
influence over the disc; when both structures are present, the radial
mixing of the stellar disc at inner radii is produced by the joint
action of the perturbers, and it is dominated by radial heating. At
outer radii, the redistribution of angular momentum is caused mostly
by the spiral arms and dominated by radial migration. This means that
the imprint of the spiral arms may be identified even in the presence
of the strong influence of the massive bar.

Similarly, the MDF in the inner region of the disc is shaped by the
action of both bar and spiral arms, while at outer radii the MDF is
mainly shaped by the spiral arms.

This study offers a new method to diagnose the large scale
non-axisymmetric structure that induces radial displacements of stars
at each radius.

The spiral arms present a characteristic signature of their existence
in observations, particularly on the radial migration and heating,
which directly relates to the shape of the MDF as seen by APOGEE.

Our approach does not consider the chemical abundance nor its
evolution, instead we employ the initial galactic position as a proxy
for metallicity. Based on the similarities of our results to the MDF
curves in the Galaxy we can state confidently that the specific
gravitational potential is of great importance to reproduce them.

\section*{Acknowledgements}
We thank the anonymous referee for a careful reading and several
excellent suggestions. We acknowledge DGTIC-UNAM for providing HPC
resources on the Cluster Supercomputer Miztli. L.A.M.M and
B.P. acknowledge DGAPA-PAPIIT through grants IN-114114 and IG-100115,
L.A.M.M and E.M. acknowledge DGAPA-PAPIIT through grant
IN105916. A.P. acknowledges DGAPA-PAPIIT through grant
IN-109716. L.A.M.M. acknowledges support from a DGAPA-UNAM
postdoctoral fellowship.


\appendix

\section{Skewness of the Initial Radii distribution at different time-frames}
\label{app:TimeFrames}

\begin{figure}
\begin{center}
\includegraphics[width=9cm]{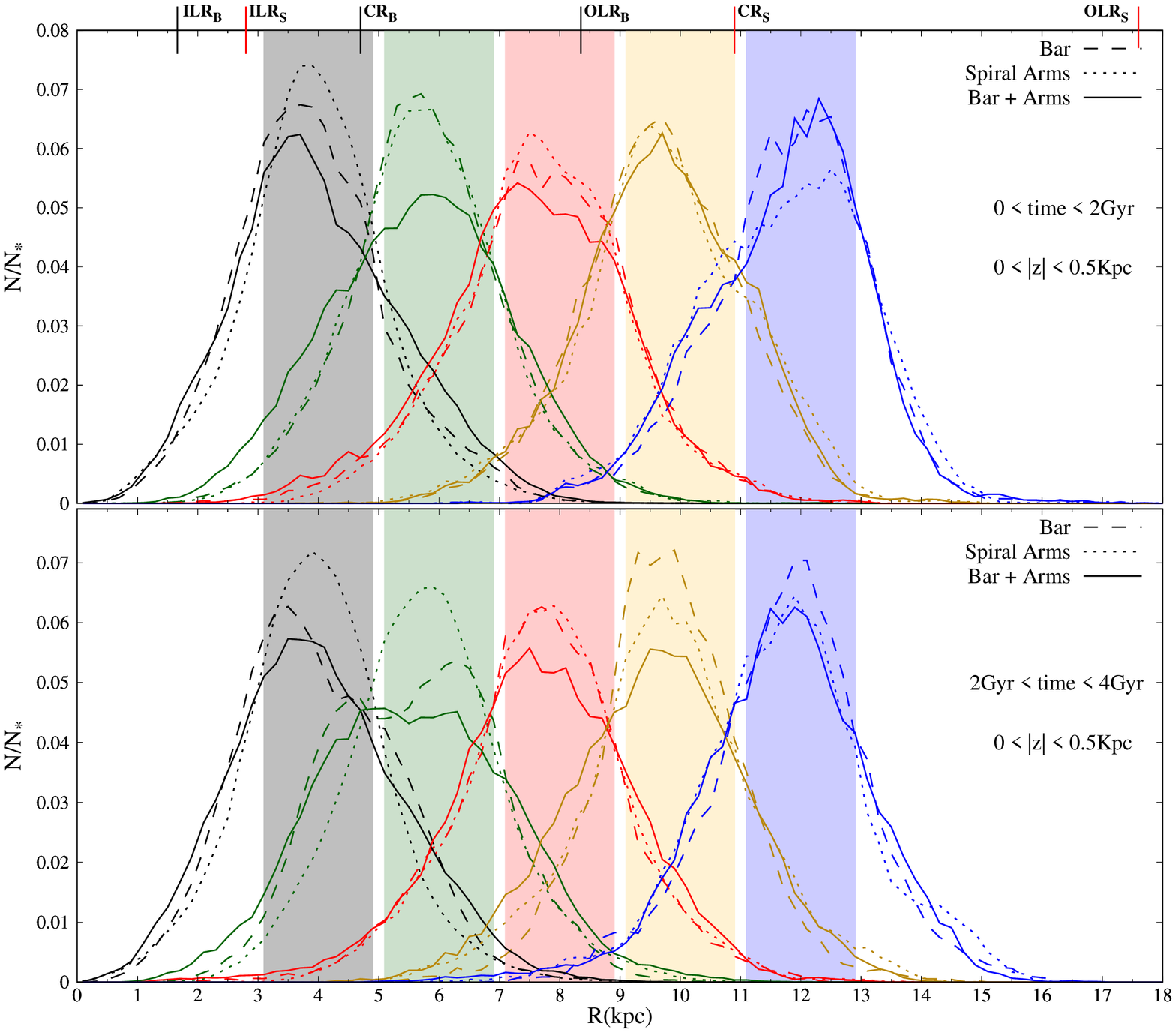}
\end{center}
\caption{Top: initial galactocentric radii distribution of stars that
  at $t_f=2$ Gyr are located within one of the five coloured
  bins. Bottom: same as top but for $t_f=4$ Gyr. In both panels, each
  curve corresponds to the closest same-colour shaded bin. $N$ is the
  number of stars at a given initial radius, and $N_*$ is the total
  number of stars that have ended within each coloured bin. Vertical
  lines at the top axis mark the positions of the ILR, CR, and the OLR
  for both the spiral pattern and the bar.}
\label{fig:Distribution2}
\end{figure}

In Figure \ref{fig:Distribution2} we show the initial radii
distributions for the time frames $0-2\Gyr$ (Top) and $2-4\Gyr$
(bottom). In both panels, the three models are still distinguishable
from each other in the same manner as was seen in Figure
\ref{fig:Distribution1} (top panel); also, notice that for this new
set of distributions the skewness keeps changing in sign when going
from inner to outer radii (as it does in Figure
\ref{fig:Distribution1}). This confirms that the results in Section
\ref{Linking} are real rather than an artefact, showing that the
spiral arms can potentially leave its imprint, through radial
migration, on the observable properties of the Milky Way's disc.

\end{document}